\documentclass{llncs}

\usepackage{listings}
\usepackage{xspace}

\newcommand{\protege}{Prot\'eg\'e\xspace}

\lstMakeShortInline[]|

\begin{document}

\title{A pattern-driven approach to biomedical ontology engineering}
\author{Jennifer D. Warrender and Phillip Lord}
\institute{School of Computing Science, Newcastle University,
  Newcastle-upon-Tyne, UK}
\maketitle

\begin{abstract}
  Developing ontologies can be expensive, time-consuming, as well as difficult
  to develop and maintain. This is especially true for more expressive and/or
  larger ontologies. Some ontologies are, however, relatively repetitive,
  reusing design patterns; building these with both generic and bespoke
  patterns should reduce duplication and increase regularity which in turn
  should impact on the cost of development.

  Here we report on the usage of patterns applied to two biomedical
  ontologies: firstly a novel ontology for karyotypes which has been
  built ground-up using a pattern based approach; and, secondly, our
  initial refactoring of the SIO ontology to make explicit use of
  patterns at development time. To enable this, we use the Tawny-OWL
  library which enables full-programmatic development of
  ontologies. We show how this approach can generate large numbers of
  classes from much simpler data structures which is highly beneficial
  within biomedical ontology engineering.
\end{abstract}

\section{Introduction}
\label{sec:introduction}

Ontologies are used widely in biomedicine for many purposes, including
instance classification, schema reconciliation, or as a controlled
vocabulary~\cite{handbook2}. The range of ontology purposes are
reflected in their sizes. Systemized Nomenclature of Medicine
(SNOMED)\footnote{\url{http://www.ihtsdo.org/snomed-ct/}} has over
290,000 concepts which presents scalability challenges when querying
the ontology. The Gene Ontology
(GO)\footnote{\url{http://www.geneontology.org/}} is about 1/10th of
this size, at 30,000 terms, while ontologies such as the Dublin Core
(DC)\footnote{\url{http://dublincore.org/}} or Simple Knowledge
Organization System
(SKOS)\footnote{\url{http://www.w3.org/TR/skos-primer/}} have less
than 100 terms each.

In some cases, the formal semantics of ontologies has been used with
computational reasoning systems. However, ontologies vary in the use
of the expressiveness of their language. Perhaps the most common is to
use just subsumption and existential restrictions (effectively the
OWL2 EL profile) of which the GO is a well-known example. Whilst SKOS
is an example of the OWL Full profile. Other ontologies use most of
the constructs within the OWL language~\cite{warren13}. While there
have historically been many ontology languages, in the life sciences
the two most common are probably OBO format~\cite{greycite11393} and
OWL~\cite{greycite2911}. Several mappings between these two
exist~\cite{tirmizi11}.

There are a variety of methodologies and approaches that exist for
ontology engineering, some of which include the use of
patterns. Patterns are templates that encourage best practice; these
were originally popularised in the context of software
engineering~\cite{gamma95}, but there has also been considerable
research on Ontology Design Patterns (ODP). ODPs are formal, reusable
and successful modelling solutions to recurrent modelling problems
that are used for creating and maintaining
ontologies~\cite{egana08}. Two well-known ODPs are:
\begin{itemize}
\item |value partition|~\cite{greycite8025}, a good practice ODP
  that is used to model attributes of objects that can only have a
  predefined set of values.
\item |sequence|~\cite{drummond06}, a domain modelling ODP. This
  ODP is used to model a sequence of events, one after the other.
\end{itemize}

There are many tools available for ontology construction, one of which
is \protege\footnote{\url{http://protege.stanford.edu/}}. Amongst
others, \protege had supported the value partition pattern through the
use of graphical
wizards\footnote{\url{http://protegewiki.stanford.edu/wiki/Protege\_Wizards}}.

As well as ODPs which address generic concerns cross-cutting a number
of domains, a need has also been recognised for patterns within a
single ontology. One mechanism for expressing these patterns is OPPL2
-- a pre-processing language, also available as a \protege plugin,
which can be used to automate addition or transformation of ontology
terms derived by a declarative rule
language~\cite{aranguren_Stevens_Antezana_2009}. A similar idea is
found with ``Safe Macros'', where patterns are expressed as annotation
properties within the ontology, which may be expanded to logical
axioms with a post-processor~\cite{mungall10}. Other tools use
patterns to leverage alternative data entry environments, generally
spreadsheets. For example, RightField~\cite{Wolstencroft15072011} and
Populous~\cite{jupp10} enable constrained data entry using an Excel
spreadsheet, and then use OPPL to expand this data into OWL
expressions. Quick term templates~\cite{roccaserra11} similarly uses
spreadsheets and the mapping language, M$^{2}$~\cite{oconnor10}.

However, ontology development using GUI based tools is time-consuming
due to the necessity of GUI interaction when there are major modelling
changes.  Therefore, a number of text based/programming environment
tools for ontology construction have been developed. Examples include
Tawny-OWL~\cite{lord_semantic_2013} (Clojure),
Thea-OWL~\cite{citeulike:6588029} (Prolog) and
InfixOWL~\cite{ogbuji08} (Python), all of which can generate
OWL. These tools enable the development of ontologies within a
programmatic environment. These also provide a mechanism for the use
of patterns; repetitive ontology construction tasks can be
automated. Next, we give a brief introduction to Tawny-OWL, and
introduce the idea of using a programmatic environment for localised
pattern development, which are (predominately) useful within a single
ontology.

\section{Patternising the Pizza Ontology}
\label{sec:tawny}

Tawny-OWL~\cite{lord_semantic_2013} is a library written in the lisp dialect
Clojure, wrapping the OWL API~\cite{horridge11}. It is designed to be used
more as a \textit{textual user interface} for ontology development, rather
than an API for ontology manipulation. Clojure provides an evaluative
environment, which means that Tawny-OWL can be used to add (or remove)
entities to an ontology incrementally and interactively. Tawny-OWL has
complete support for OWL2. including data types. It also provides direct
access to reasoners, which combined with unit testing and a versioning system
enables continuous integration of ontologies. Tawny-OWL has been designed to
be convenient for building simple ontologies without in-depth knowledge of
Clojure, however, fully programmatic use of Tawny-OWL will require this
knowledge.

The syntax of Tawny-OWL is relatively straight-forward, having been
modelled after Manchester Syntax~\cite{greycite2216}. Consider, for
example, this definition (Listing~\ref{lst:class}) from the Pizza
Ontology\footnote{\url{http://owl.cs.manchester.ac.uk/co-ode-files/ontologies/pizza.owl}}
recast into Tawny-OWL syntax (aka tawny-pizza)\footnote{Available from
  \url{https://github.com/phillord/tawny-pizza}}, which we use here as
an exemplar of Tawny-OWL and patterns.

\begin{lstlisting}[caption={A basic class definition.},label=lst:class]
(defclass Pizza
   :label "Pizza"
   :subclass (owl-some hasTopping PizzaTopping)
   (owl-some hasBase PizzaBase))
\end{lstlisting}

As Tawny-OWL is built on a full programming language it is also
capable of expressing arbitrarily complex patterns. It provides
support for two design patterns which are so commonly used that it
they are rarely recognised as a pattern, the first being the closure
axiom~\cite{greycite1296}. In Listing~\ref{lst:closed}, we show an
expression which returns the two existential, and one universal
restrictions necessary to describe the toppings for a Margherita
pizza. Covering axioms are also supported.

\begin{lstlisting}[caption={A closed restriction.},label=lst:closed]
;; A - Usage
(some-only hasTopping TomatoTopping MozzarellaTopping)

;; B - As Manchester Syntax
hasTopping some TomatoTopping
hasTopping some MozzarellaTopping
hasTopping only 
   (MozzarellaTopping or TomatoTopping)
\end{lstlisting}

Currently, Tawny-OWL has preliminary support for other general purpose
ontology design patterns, in the form of the value partition ODP. The
usage of this Tawny-OWL pattern can be seen in
Listing~\ref{lst:valuepartition}, which generates four classes, a
disjoint axiom and an object property.

\begin{lstlisting}[caption={Example value partition usage in the Pizza
    Ontology.},label=lst:valuepartition]
(value-partition
 Spiciness
 [Mild Medium Hot])
\end{lstlisting}

Closure axioms and the value partition are currently the only generic
patterns used in the Pizza ontology. We do use a localised pattern,
which we called ``named pizza'' where a particular pizza is defined by
an enumeration of its ingredients (Listing~\ref{lst:namedpizza}). This
pattern also makes use of the closure pattern described earlier.

\begin{lstlisting}[caption={Example of localised patterning in the
    Pizza Ontology. In this example, \texttt{owl-class}, and
    \texttt{some-only} are parts of OWL, \texttt{hasTopping} is an
    object property, \texttt{pizzalist}, \texttt{named} and
    \texttt{toppings} are variables, while \texttt{defn} and
    \texttt{doseq} are parts of Clojure.},label=lst:namedpizza]
(defn generate-named-pizza [& pizzalist]
  (doseq [[named & toppings] pizzalist]
    (owl-class
     named
     :subclass NamedPizza
               (some-only hasTopping toppings))))
\end{lstlisting}

The syntactic concision of this pattern, as shown in
Listing~\ref{lst:namedexamples}, is advantageous as a pizza can have
many toppings. A secondary advantage is to ensure consistency of all
the named pizza definitions, as well as supporting maintainability
should we wish to change these definitions. Tawny-pizza is currently
an example usage -- in real usage, we would probably read this data
from a spreadsheet, or a simpler flat file, similar to tools such as
RightField.

\begin{lstlisting}[caption={Example Named Pizza
    inputs.},label=lst:namedexamples]
 [CajunPizza MozzarellaTopping OnionTopping PeperonataTopping
  PrawnsTopping TobascoPepperSauce TomatoTopping]

 [CapricciosaPizza AnchoviesTopping MozzarellaTopping
  TomatoTopping PeperonataTopping HamTopping CaperTopping
  OliveTopping]
\end{lstlisting}

In this section, we have introduced Tawny-OWL and shown how it enables
the application of patterns; however, this has been in the context of
the pizza ontology which is only an exemplar ontology, rather than one
intended for real use. In the next section we consider the use of
patterns in the Karyotype Ontology.

\section{Patternising a novel ontology}
\label{sec:karyotype}

Next we are going to describe the usage of patterns applied to a novel
ontology for karyotypes which has been built ground-up. Here, we show
that the ontology would be difficult to write by hand, and therefore
that ontology construction is aided by Tawny-OWL and the use of
patterns. The Clojure code for the Karyotype project is available at
\url{https://github.com/jaydchan/tawny-karyotype}.

First, we introduce \textit{karyotypes} which are a description of all
the chromosomes in a cell. Humans have 46 chromosomes, in 23
pairs. Locations along the chromosome can be identified by their
patterns of \textit{chromosome bands} which are visible under light
microscopy. A karyotype describes the total number of chromosomes, the
sex chromosomes and any chromosomal abnormalities (if
any). Abnormalities are described by their kind (e.g.  insertion,
deletion) and the location they affect defined relative to the visible
bands. Human karyotypes are normally represented in string format, as
defined by the International System for human Cytogenetic Nomenclature
2009 (ISCN2009)~\cite{shaffer09}. For example, the karyotype |46,XY|
is a normal male karyotype. However, current ISCN strings can be
complicated, lack formal interpretation, and are not computationally
amenable (trivially, they cannot even be represented in ASCII as they
include meaningful underlining, used to distinguish homologous
chromosomes). As a result, these ISCN strings can be hard to parse,
validate and query. This is also true for the ISCN specification.

In order to overcome these problems, we have developed the Karyotype
Ontology~\cite{warrender13}. For this work, we made use of a pragmatic
methodology to model karyotypic information, meaning we made no
distinctions unless required by the use cases identified in the
ISCN2009. The karyotype ontology demonstrates the requirement for a
pattern driven approach, as it is highly repetitive; almost all of the
entities are part of at least one pattern.

Our initial construction of the Karyotype Ontology involved the use of
a partonomic structure, built in Manchester Syntax. A resulting
definition of human chromosome 1 can be seen in
Listing~\ref{lst:priorhuman}. This ontology, however rapidly became
unmanageable simply because of the number of restrictions on each
chromosome. This is made more complex still because banding patterns
are visible at 5 different resolutions, each containing more bands
than the last.

\begin{lstlisting}[caption={Incomplete chromosome definition for a
    normal Human Chromosome 1 using strict
    partonomy.},label=lst:priorhuman]
Class: NormalHumanChromosome1
   SubClassOf:
      HumanAutosome
      hasPart exactly 1 HumanChromosomeBand1pTer
      hasPart exactly 1 HumanChromosomeBand1p36.3
      hasPart exactly 1 HumanChromosomeBand1p36.2
      ... (23 other hasPart relations removed)
      hasPart exactly 1 HumanChromosomeBand1qTer
\end{lstlisting}

The use of Tawny-OWL enables these concepts to be generated to a standard
pattern. Listing~\ref{lst:bandusage} shows an expression using the
|humanbands| pattern, for a (small!) subset of the human chromosome bands.
Additionally, we have specialised and inverted the |hasPart| relationship to
|isBandOf| and |isSubBandOf|, as this produces smaller classes with fewer
restrictions, which is simpler to work with\footnote{At the current time, we
  are uncertain whether this small change in semantics will impact the
  reasoning we may wish to perform. If we need to invert, or add both the
  forward and inverse relationship at a later date, the programmatic nature of
  Tawny-OWL makes this easy to achieve}. The OWL ontology which is generated
as a result is shown in Listing~\ref{lst:bandoutput} as Manchester Syntax.

\begin{lstlisting}[caption={Incomplete chromosome definition for a
    Human Chromosome 1 showing the input format of the human bands
    pattern.},label=lst:bandusage]
(humanbands HumanChromosome1
 ["p36.3" "p36.33" "p36.32" "p36.31"])
\end{lstlisting}

\begin{lstlisting}[caption={Elided output from
    Listing~\ref{lst:bandusage}.}, label=lst:bandoutput]
;; A - Generic chromosome 1 band definitions
Class: HumanChromosome1Band
   SubClassOf:
      HumanChromosomeBand
      isBandOf some HumanChromosome1

Class: HumanChromosome1Bandp
   SubClassOf:
      HumanChromosome1Band

;; B - Chromosome 1 band definition
Class: HumanChromosomeBand1p36.3
   SubClassOf:
      HumanChromosome1Bandp

;; C - Chromosome 1 sub-band definition
Class: HumanChromosomeBand1p36.31
   SubClassOf:
      HumanChromosome1Bandp
      isSubBandOf some HumanChromosomeBand1p36.3
\end{lstlisting}

The bulk of the karyotypes ontology consists of entities which are
part of this pattern, the total size of which is shown in
Table~\ref{tab:humanstats}.  In practice, building this ontology
manually would have been impractical, particularly from a
maintainability point of view. Therefore, to aid construction we
tested tools such as OPPL and Populous. However, both had difficulties
because of the split between different source files and forms for
expressing the patternised and non-patternised sections of the
ontology. This contrasts with Tawny-OWL which uses a single syntax,
and also provides testing and reasoning services again within a single
syntax.

\begin{table}
  \begin{center}
    \caption{Human Chromosome model statistics.}
    \label{tab:humanstats}
    \begin{tabular}{lll}
      \hline\noalign{\smallskip}
      Class Type & Biological Object & Number of Classes\\
      \noalign{\smallskip}
      \hline
      \noalign{\smallskip}
      Chromosome & 24 & 27\\
      Centromere & 24 & 25\\
      Telomere & 48 & 25\\
      Bands and Sub-bands & 1213 & 1286\\
      \noalign{\smallskip}
      \hline
      \noalign{\smallskip}
      Total Number & 1309 & 1363\\
      \hline
    \end{tabular}
  \end{center}
\end{table}

The |humanbands| pattern is not the only pattern in KO. While
canonical chromosomes are expressed as a partonomy, this is not
sufficient to express all abnormal karyotypes. As a simple example, a
karyotype can be defined by its loss of a chromosome, as opposed to
its congenital absence: the karyotype |45,X,-Y| is a male karyotype
where the Y chromosome has been lost; although partonomically
identical to the karyotype |45,X|, it is considered different as the
latter is a congenital absence.

Therefore, karyotypes are expressed as events, similar to the way ISCN
strings are modelled. Using the event-based change approach, a |45,X|
female is described as |46,XN|, with a deletion of a sex chromosome
(see Listing~\ref{lst:45X}), in this case N represents an unknown sex
chromosome, as opposed to |45,X,-Y| which is derived from a |46,XY|
male.

\begin{lstlisting}[caption={The karyotypic definition for
    45,X.},label=lst:45X]
(defclass k45_X
  :label "The 45,X karyotype"
  :subclass ISCNExampleKaryotype
  (owl-some derivedFrom k46_XN)
  (deletion 1 HumanSexChromosome))
\end{lstlisting}

Within the Karyotype Ontology, we define an event as a concept,
supported by a usage pattern and a function which generates a
restriction according to this pattern. One simple example is the
inverse event pattern and its usage is shown in
Listing~\ref{lst:inversion}.

\begin{lstlisting}[caption={The pattern, usage and resultant OWL used
    to define inverse events.},label=lst:inversion]
;; A - Inversion pattern
(defn inversion [n band1 band2]
    (exactly n hasEvent 
        (owl-and Inversion 
            (owl-some hasBreakPoint 
                      band1 band2))))

;; B - Usage
(inversion 1 2p21 2q31)

;; C - As Manchester Syntax
hasEvent exactly 1
         (Inversion and (hasBreakPoint some 2p21 2q31))
\end{lstlisting}

While the large, partonomic section of the karyotype is now complete, the
ontology is still being developed. In time, we intend to extend the ontology
and expect that the generators will be become the main ``user interface'' to
the work, in order to present an end-user syntax. Patterns have, therefore,
been useful in the development of the Karyotype Ontology, and will be so in
downstream usage.

\section{Patternised development of an existing ontology}
\label{sec:sio}

While the development of the Karyotype Ontology shows that patterns
have been useful in this one context, this does not demonstrate that
it is generally useful; therefore, in this section, we apply a similar
development methodology to an existing ontology, namely SIO. We show
that, while in the Karyotype Ontology we see a few patterns generating
many entities, in SIO we see the reverse; there are many patterns
generating relatively few entities in the ontology construction. As
with Karyotype Ontology, we can exploit patterns in downstream usage.

Semanticscience Integrated Ontology (SIO)~\cite{greycite11363} is a
simple upper level OWL ontology for the integration of types and
relations that provides rich descriptions of objects, processes and
their attributes. It defines 1395 classes, 202 object properties, 1
data property and 8 annotation properties~\cite{dumontier13}. We chose
SIO as it is explicit in promoting the use of ODPs to describe and
associate numerous entities e.g. qualities and capabilities. In this
paper, \textit{SIO} is used to refer to the existing ontology, whilst
\textit{tawny-sio} refers to the Tawny-OWL recasting of SIO. The
Clojure code for tawny-sio project is available at
\url{https://github.com/jaydchan/tawny-sio}.

SIO has not been developed in Tawny-OWL and therefore is only
available as an OWL file. To enable the development of a patternised
form of SIO, first rendered this OWL file into Tawny-OWL syntax. SIO
uses numeric IDs as the fragment of its URL. While there are good
reasons for this, it means that the fragment is unsuitable at a code
level as a memorable identifier for the entity. Therefore Tawny-OWL
provides a Clojure-safe syntactic transformation of the
rdfs:label. Hence, SIO\_000395 becomes |to_regulate|, with a few
specific replacements for tawny-sio concepts that transform to
reserved words (e.g. ``true'' and ``false'').

We now describe how patterns have been applied to tawny-sio. A generic
pattern was identified that is useful for most of the tawny-sio
classes (Listing~\ref{lst:sioclass}), which supports a SIO standard
that (almost) all classes have a name, parent and textual
description. The |sio-class| enforces this, as well as providing
syntactic sugar.

\begin{lstlisting}[caption={A common pattern for tawny-sio
    classes. \texttt{name}, \texttt{parent}, \texttt{description} and
    \texttt{frames} are variables.},label=lst:sioclass]
(defn sio-class [name parent description & frames]
  (apply owl-class
         (list* (make-safe name)
                :subclass parent
                :label name
                :annotation (desc description)
                frames)))
\end{lstlisting}

The |sio-class| function makes use of a second pattern, namely the description
pattern which adds a standardized annotation using the Dublin Core
ontology, as show in Listing~\ref{lst:description}.

\begin{lstlisting}[caption={The description pattern for tawny-sio
    classes.},label=lst:description]
(def dc-description (iri "http://purl.org/dc/terms/description"))
(defn desc [description]
   (annotation dc-description
       (literal description :lang "en")))
\end{lstlisting}

The exceptions to the |sio-class| pattern, are the concepts that model
the chemical elements (atoms). An alternative pattern is used for
these; the encoding of this pattern is shown in
Listing~\ref{lst:owlatom}. As this pattern is specialised for atoms,
the superclass is ``hard-coded'' into the pattern.

\begin{lstlisting}[caption={The atom generator
    function. \texttt{owl-atom} is used to distinguish from
    \texttt{atom} which is used by Clojure.},label=lst:owlatom]
;; A - See-also pattern
(defn see-also [value]
   (annotation seeAlso
       (literal value :type :RDF_PLAIN_LITERAL)))

;; B - Auxiliary function
(defn owl-atom-annotation-maybe [cls chebi]
  (if-not (nil? chebi)
   (add-annotation
    cls (see-also chebi))))

;; C - Atom pattern
(defn owl-atom [name chebi]
 (owl-atom-annotation-maybe
  (owl-class (make-safe name)
             :subclass atom
             :label name)
  chebi))
\end{lstlisting}

One interesting outcome of encoding this pattern concerns
ChEBI~\cite{Hastings01012013} annotations. Most SIO elements have a
|seeAlso| annotation that links the atom to its equivalent ChEBI
ID. However, elements 112 to 118 lack this annotation, although 112
aka Copernicium aka ununubium can be found with ChEBI value
CHEBI:33517. Encoding this pattern, forces us to
deal with these exceptions explicitly.

\section{Patterns for downstream usage}
\label{sec:patt-downstr-usage}

The SIO wiki pages includes many exemplar ODPs that can be used in
conjunction with SIO. One such ODP is the biochemical pathway
pattern. This ODP includes a variation of the sequence
ODP~\cite{drummond06}. The biochemical pathway ODP (see
Listing~\ref{lst:biologicalpathway}), as well as other ODPs, has been
encoded into tawny-sio and available for downstream usage.

\begin{lstlisting}[caption={The Clojure function and auxiliary
    function for biochemical pathway pattern, its usage and resulting
    OWL class. \texttt{name} and \texttt{reactions} are
    variables.},label=lst:biologicalpathway]
;; A - Precedes pattern
(defn biochemical-pathway0 [reactions]
  (owl-and (first reactions)
           (owl-some precedes (rest reactions))))

;; B - Biological pathway pattern
(defn biochemical-pathway [name reactions]
  (owl-class name
            :equivalent
            (owl-and pathway
                     (owl-some has_proper_part
                               (biochemical-pathway0 reactions))
                     (owl-some has_proper_part reactions))))

;; C - Usage
(biochemical-pathway "glycosis"
                     [hexokinase_reaction
                      phosphoglucose_isomerase_reaction
                      ...])

;; D - As Manchester Syntax
Class: 'glycolysis'
   EquivalentTo:
      'pathway'
       and 'has proper part' some
           ('hexokinase reaction' and 'precedes' some
           ('phosphoglucose isomerase reaction' and 'precedes' ...)
       and 'has proper part' some 'hexokinase reaction'
       and 'has proper part' some 'phosphoglucose isomer reaction'
       ...
\end{lstlisting}

The |biochemical-pathway| generator function is not actually used as
part of tawny-sio; SIO is intended for use as an upper and middle
ontology, and does not, therefore, model any pathways itself. Instead
SIO documents the pattern for downstream users. However, there is no
computational representation of this pattern. Within tawny-sio, we can
provide such a representation which is not only descriptive but which
can be used to computationally generate specific pathways. As we have
also described with the Karyotype Ontology, this pattern becomes a
part of the ``user interface'' of the ontology. As well as this
pattern, we now have generators for molecule, enzyme and biochemical
reaction patterns.

As with the atoms, encoding these exemplar ODPs has highlighted some
interesting issues. For example, the |target_role| class associated
with the biochemical pathway
ODP\footnote{\url{http://code.google.com/p/semanticscience/wiki/ODPBiochemistry}}
is missing from SIO. Further investigation of the missing
\texttt{target\_role} class, shows that the actual role class is
\texttt{reactant\_role}~\cite{boelling12}. The process of the
computable encoding of the patterns, is in itself a useful process for
quality control and consistency of the ontology.

In this section, we have shown how suggested usage patterns for an
ontology can become part of the ontology.

\section{Discussion}
\label{sec:discussion}

In this paper, we discuss the use of generalised and localised
patterns in ontology engineering. We describe the use of patterns in
three distinct ontologies: tawny-pizza, an \textit{exemplar} recasting
of the pizza ontology into Tawny-OWL; the Karyotype Ontology, a
\textit{novel} ontology built for describing karyotypes using an
event-based approach; and tawny-sio, a fork of an \textit{existing}
upper ontology describing scientific objects, processes and their
attributes. We demonstrate that for the Karyotype Ontology, most
entities are part of one or more localised patterns. For SIO, the
localised patterns encourage consistency and can be made available for
downstream users.

There are other tools for ODPs such as OPPL2 and safe macros. However,
Tawny-OWL has the advantage that a single syntax is used for both the
patternised and non-patternised parts of the ontology. The basic
syntax of Tawny-OWL is similar to Manchester Syntax, is
``unprogrammatic'' and should be usable by non-programmers, without
significant, additional effort, especially when compared to the
equivalent Java code.

Tawny-OWL currently provides support for three well-known existing
patterns: value partition, closure and covering axioms. These patterns
have been used in tawny-pizza (see Section~\ref{sec:tawny}) and are
available for use in other ontologies. However, as it is fully
programmatic, Tawny-OWL can encode patterns, localised to the scope of
a single ontology. These localised patterns are useful as they enable a
concise syntax, ensure consistency and support maintainability should
the need for change arise. The Karyotype Ontology is predominantly
made up of a variety of localised patterns -- examples include the
|humanbands| pattern and |inversion| pattern (see
Section~\ref{sec:karyotype}). Currently the ontology is still being
developed and our patterns are incomplete. For example, karyotypic
events such as deletions and insertions affect a sequence of bands,
which are currently not modelled in our ontology. There are 3 ways we
could achieve this: firstly to utilise a variant of the sequence
design pattern~\cite{drummond06}; secondly the assignation of ordinal
numbers to the chromosome bands as a datatype; and, finally, we could
use a pattern in Clojure which expands to all the affected
bands. \textit{A priori}, it is difficult to determine which of these
will work best, particularly with respect to non-functional
characteristics such as reasoning time. The use of Tawny-OWL will
enable us to test this by generating multiple test versions of the
ontology. With 800 classes this would otherwise be impractical.

We have demonstrated that patterns have benefited in both tawny-pizza
and in the construction of the Karyotype Ontology. However, we wish to
understand whether patterns are generally useful. Therefore, we have
also applied this methodology to SIO, which we chose after the
construction of Tawny-OWL, and which was built without knowledge of
Tawny-OWL.

We find that patterns are less useful within tawny-sio than the Karyotype
Ontology. However, a number of patterns were identified and their use could
increase the consistency and concision of this ontology. Furthermore, there
are additional patterns, which whilst not themselves used in tawny-sio, are
potentially useful for downstream users of tawny-sio.

The utility of the pattern approach will depend on the nature of the
ontology.  For example, a structurally simple ontology may use few
patterns. However, we note that patterns are not limited to the
logical component of OWL; within tawny-sio we have used a number of
annotation patterns.

While the use of general ontology design patterns is well documented,
the use of localised patterns is less so. In this paper, we have
described the application of Tawny-OWL to three ontologies which have
allowed us to test the utility of this form of pattern. It appears to
be a promising methodology which could substantially impact ontology
engineering.

\bibliographystyle{splncs}
\bibliography{mybib}

\end{document}